\documentclass[11pt,a4paper]{article}
\usepackage[margin=2.4cm]{geometry}
\usepackage{amsmath,amssymb,bm}
\usepackage[colorlinks=true,linkcolor=blue,citecolor=blue,urlcolor=blue]{hyperref}
\usepackage{booktabs}
\newcommand{\tth}{\tilde{\theta}}
\newcommand{\tQ}{\tilde{Q}}
\newcommand{\tD}{\tilde{D}}
\newcommand{\bD}{\bar{D}}
\newcommand{\btD}{\bar{\tilde{D}}}
\newcommand{\eps}{\varepsilon}
\newcommand{\half}{\tfrac{1}{2}}

\title{\bf Kalb--Ramond Topological Term in Majorana Superspace and\\ Kaluza--Klein Spectrum Deformation in Five Dimensions}
\author{L.~A.~S.~Nunes$^{1}$\thanks{luciana@ufersa.edu.br} \ and \ C.~A.~S.~Almeida$^{2}$\thanks{carlos@fisica.ufc.br}\\[4pt]
{\normalsize\it $^{1}$Centro de Ci\^encias Exatas e Naturais, Universidade Federal Rural do Semi-\'Arido (UFERSA),}\\
{\normalsize\it Mossor\'o-RN, 59625-900, Brazil}\\
{\normalsize\it $^{2}$Universidade Federal do Cear\'a, Departamento de F\'isica, 60455-760, Fortaleza-CE, Brazil}}
\date{}

\begin{document}
\maketitle

\begin{abstract}
\noindent
We construct an intrinsic $N=1$, $D=5$ superspace whose Grassmann coordinate is a five-dimensional Dirac spinor decomposed into two four-dimensional Majorana spinors, $\Theta=\theta+i\tth$, and use it to analyze the supersymmetric structure of the five-dimensional Kalb--Ramond (KR) topological term. The supersymmetry algebra closes on the fifth momentum through anticommutators of \emph{opposite-sector} generators, realizing $\partial_5$ as a central charge in the sense of Sohnius; we show that this assignment is the unique covariant one, and we give an explicit dictionary to the $N=1/2$ superspace of Linch, Luty and Phillips and to the four-dimensional $N=1$ superform hierarchies of Becker, Becker, Linch, Randall and Robbins. Two structural results follow. First, a tensor multiplet subjected to both available same-side chirality conditions cannot propagate in the fifth dimension; the KR multiplet is therefore described by a singly chiral spinor superfield whose composite field strength $T_{5\alpha}=\partial_5 B_\alpha+W_\alpha(V_T)$ realizes the single-chain tensor hierarchy with differential $\partial_5$, and the supersymmetric completion of the topological term is the corresponding $BF$-type quadratic invariant, whose component form agrees, up to normalization conventions, with the pseudo-supersymmetric action. Second, we correct the six-dimensional origin of the topological term, $S_6=\int d\phi\wedge B\wedge H$, and show that for the KR field alone the five-dimensional term is a total derivative and leaves the Kaluza--Klein tower undeformed. After the superfield-level identification $B_{m5}\to A_m$, however, the mixed topological term becomes a genuine $BF$ coupling between independent multiplets, and the Kaluza--Klein spectrum splits exactly into $m_{n,\pm}^2=\big(\sqrt{n^2/R^2+M^2}\pm M\big)^2$, with $M$ fixed by the topological coupling: the zero mode acquires a topological mass $2M$, and the excited towers are displaced by the constant $\pm M$ relative to bulk fields that share the compactification radius but lack the coupling. This relative deformation cannot be absorbed into a redefinition of $R$ and constitutes a sharp, calculable prediction for compactification on $S^1$. On the orbifold $S^1/\mathbb{Z}_2$ relevant to brane worlds the situation is qualitatively different, and we show that it is controlled by supersymmetry: the gauge structure of the hierarchy correlates the $\mathbb{Z}_2$ parities of the two multiplets, the topological term is parity-odd, and a constant coupling is projected out, so that the deformation is governed by a coupling of the kink type --- precisely the one generated by the six-dimensional parent action --- and becomes second order in the coupling, with brane-localized contributions at the fixed points.
\end{abstract}

\section{Introduction}
\label{sec:intro}

The interest in higher-dimensional field theories has been greatly stimulated by the Randall--Sundrum (RS) brane-world scenario \cite{RS1,RS2}, which provides a geometrical mechanism for addressing the hierarchy problem without large extra dimensions. The observable universe is identified with a four-dimensional hypersurface (brane) embedded in a five-dimensional anti-de Sitter bulk, with the Standard Model fields localized on the brane while gravity propagates through the extra dimension. A natural consequence of the scenario, inherited from its string-theoretic motivation, is the presence of a rank-2 antisymmetric tensor -- the Kalb--Ramond (KR) field \cite{KR} -- a massless excitation of the closed-string spectrum that couples to the world-sheet.

The KR field is central to the generation of spacetime torsion: when a five-dimensional Einstein--Cartan theory is supplemented by a bulk KR field, the rank-3 field strength $\mathcal H_{MNP}$ sources the torsion, and its zero mode couples to brane fields with a strength exponentially suppressed by the warp factor \cite{Mukho1,Mukho2}. This suppression has been proposed as an explanation for the low-energy invisibility of torsion, while massive KR modes may leave imprints at TeV-scale experiments.

In five dimensions the KR field supports a Chern--Simons-like coupling,
\begin{equation}
S_{\rm top}=k\int d^5x\,\eps^{MNPQR}\,B_{MN}\mathcal H_{PQR},
\label{eq:Stop-intro}
\end{equation}
which, as we discuss in Section~\ref{sec:top}, descends from the six-dimensional term $\int d\phi\wedge B\wedge H$ with a scalar profile $\phi(z)$ localized along the sixth direction, and which generalizes the axion--photon coupling behind the Peccei--Quinn quasi-symmetry \cite{PQ,TahimAlmeida}. Being built entirely from forms, the coupling is metric-independent, a desirable feature in brane-world settings.

Supersymmetry enters this problem not as a proposed solution to the hierarchy problem, but as a structural organizing principle. Independently of its phenomenological status, supersymmetry is the unique consistent extension of the Poincar\'e algebra by fermionic generators \cite{HLS}, it is demanded by the string-theoretic origin of the bulk fields, and -- most relevant here -- the supersymmetric completion of a given bulk term is a sharply posed question. In five dimensions, $N=1$ supersymmetry carries eight supercharges, and a manifestly supersymmetric formulation requires a superspace adapted to the $D=5$ geometry.

Two complementary strategies exist. The first employs harmonic or projective superspace, with all eight supercharges manifest; this is the approach of Kuzenko and collaborators \cite{KuzenkoNovak,KuzenkoLinch}. The second represents four supercharges manifestly and the remaining four linearly, in a 4D, $N=1$ superspace augmented by a fifth bosonic coordinate $y=x^5$ -- the $N=1/2$ approach of Linch, Luty and Phillips \cite{LLP}, developed at the supergravity level in \cite{BBBLR}. Within the second strategy, the general theory of $p$-form multiplets and of their Chern--Simons-like couplings has been developed by Becker, Becker, Linch, Randall and Robbins as the superspace embedding of Abelian tensor hierarchies \cite{TAMU1,TAMU2}: the extra-dimensional derivative enters as the differential of a chain complex, and all quadratic and cubic invariants have been classified.

We adopt a variant of the second strategy: an intrinsic $N=1$, $D=5$ superspace in which the Grassmann coordinate $\Theta$ is a full five-dimensional Dirac spinor, decomposed as $\Theta=\theta+i\tth$ into two four-dimensional Majorana spinors. The distinguishing feature of this basis is that the fifth momentum appears directly in the supersymmetry algebra, as a central charge in the sense of Sohnius \cite{Sohnius}: anticommutators of generators from \emph{opposite} Majorana sectors close on $\partial_5$, while same-sector anticommutators close on the four-dimensional momentum. We show that this assignment is not a choice but a theorem: since $C^{-1}\gamma^5$ is antisymmetric in four dimensions, a $\partial_5$ term attached to the same-sector Grassmann coordinate can never be covariantized. The Majorana basis is particularly convenient for brane-world applications: the orbifold parity acts simply as $\tth\to-\tth$, and bulk Dirac fermions couple without any change of spinor basis.

Within this framework we obtain the following results.
\begin{itemize}
\item[(i)] We construct the superspace generators and covariant derivatives and establish the unique covariant form of the algebra, with $\partial_5$ realized as a central charge (Section~\ref{sec:superspace}). An explicit dictionary relates the construction to the $N=1/2$ formalism and to the pseudo(symplectic)-Majorana superspace of \cite{KuzenkoLinch}.
\item[(ii)] We prove a structural no-go result: a superfield subjected to both same-side chirality conditions, $\bD\Phi=\btD\Phi=0$, necessarily obeys $\partial_5\Phi=0$. The KR tensor multiplet must therefore be described by a singly chiral spinor superfield $B_\alpha$, exactly as in the $N=1/2$ framework (Section~\ref{sec:tensor}).
\item[(iii)] The composite field strength $T_{5\alpha}=\partial_5B_\alpha+W_\alpha(V_T)$ is recognized as the single-chain case of the tensor-hierarchy field strength of \cite{TAMU1,TAMU2}, with the abstract differential realized as $q\to\partial_5$. The supersymmetric completion of the topological term is the corresponding $BF$-type quadratic invariant; its component expansion reproduces the pseudo-supersymmetric action of Klein \cite{Klein} up to normalization conventions, with no additional structures, and the completeness of the classification of \cite{TAMU2} guarantees that no further quadratic invariants exist. The fermionic sector of the invariant is itself topological (Section~\ref{sec:top}).
\item[(iv)] We correct the six-dimensional origin of the topological term: the parent action is $S_6=\int d\phi\wedge B\wedge H$, which localizes to \eqref{eq:Stop-intro} on slices where the gradient of $\phi$ is supported. For the KR field alone, \eqref{eq:Stop-intro} is the total derivative $\half\,d(B\wedge B)$ and leaves the Kaluza--Klein spectrum undeformed. After the superfield-level identification $B_{m5}\to A_m$, however, the mixed topological term becomes a genuine $BF$ coupling between independent multiplets, and the spectrum splits exactly into
\begin{equation}
m_{n,\pm}^{2}=\Big(\sqrt{n^{2}/R^{2}+M^{2}}\pm M\Big)^{2},
\label{eq:spectrum-intro}
\end{equation}
with $M$ fixed by the topological coupling: topological mass generation at the zero mode, and a constant, non-absorbable displacement $\pm M$ of the excited towers relative to bulk fields lacking the coupling (Section~\ref{sec:KK} and Appendix~\ref{app:KK}). Section~\ref{sec:orbifold} establishes the structure of the $S^1/\mathbb{Z}_2$ projection, where the topological term is $\mathbb{Z}_2$-odd and the deformation becomes second order in the coupling.
\end{itemize}

The paper is organized as follows. Section~\ref{sec:superspace} collects the $D=5$ Clifford-algebra conventions and constructs the intrinsic superspace, emphasizing its relation to the $N=1/2$ formalism. Section~\ref{sec:tensor} presents the tensor supermultiplet, derives its component expansion and the invariant action, and relates the construction to the superform hierarchies. Section~\ref{sec:top} discusses the topological term, its six-dimensional origin and the identification $B_{m5}\to A_m$; Section~\ref{sec:KK} derives the deformed Kaluza--Klein spectrum. Section~\ref{sec:conclusions} contains our conclusions. Appendix~\ref{app:dof} reviews the degree-of-freedom counting behind the identification, Appendix~\ref{app:KK} details the mode diagonalization, and Appendix~\ref{app:nogo} records the verification that the covariant derivative structure generates no component terms beyond those displayed in the text.

\section{The intrinsic $N=1$, $D=5$ superspace}
\label{sec:superspace}

\subsection{Clifford algebra and spinor conventions}
\label{sec:conventions}

We work in five-dimensional Minkowski space with metric signature $(+,-,-,-,-)$. Capital Latin indices $M,N,P,\dots$ run over $\{0,1,2,3,5\}$ and lower-case Latin indices $m,n,k,l,\dots$ run over the four-dimensional subspace $\{0,1,2,3\}$; this single convention is used throughout the paper and the appendices, and Greek letters are not used as spacetime indices. The Levi-Civita symbol is normalized by $\eps^{01235}=+1$. Antisymmetrization is taken with unit weight, so that the five-dimensional Kalb--Ramond field strength is
\begin{equation}
\mathcal H_{MNP}=\partial_{[M}B_{NP]}=\partial_MB_{NP}+\partial_NB_{PM}+\partial_PB_{MN},
\label{eq:Hdef}
\end{equation}
with four-dimensional and mixed components $\mathcal H_{mnk}$ and $\mathcal H_{mn5}$. We use the symbol $\mathcal H$ exclusively for this object; the differential-form notation $H=dB$ of Section~\ref{sec:bosonic} refers to the same field strength, $\mathcal H_{MNP}$ being its components.

The five-dimensional Clifford algebra $\{\Gamma^M,\Gamma^N\}=2\eta^{MN}$ is realized as
\begin{equation}
\Gamma^m=\gamma^m=\begin{pmatrix}0&\sigma^m\\ \bar\sigma^m&0\end{pmatrix},
\qquad
\Gamma^5=i\gamma^5,\qquad \gamma^5=i\gamma^0\gamma^1\gamma^2\gamma^3 ,
\label{eq:cliff}
\end{equation}
with $\sigma^m=(\mathbf 1,\vec\sigma)$, $\bar\sigma^m=(\mathbf 1,-\vec\sigma)$. The charge-conjugation matrix is
\begin{equation}
C=i\gamma^2\gamma^0,\qquad C\gamma^m C^{-1}=-(\gamma^m)^T,\qquad C^T=-C,\qquad C^2=-\mathbf 1 ,
\label{eq:C}
\end{equation}
and the Majorana conjugate is $\bar\psi=\psi^TC$. Two Lorentz generators occur below and must be distinguished, since they differ in size and in normalization:
\begin{equation}
\sigma^{mn}=\tfrac14\big(\sigma^m\bar\sigma^n-\sigma^n\bar\sigma^m\big)\quad(2\times2),
\qquad
\Sigma^{mn}=\tfrac14\big[\gamma^m,\gamma^n\big]\quad(4\times4).
\label{eq:sigmadef}
\end{equation}
The two-component $\sigma^{mn}$ is the one appearing in the Weyl-basis component expansions of Section~\ref{sec:WZ}; the four-component $\Sigma^{mn}$ is the one appearing in the symmetry table below and in the five-dimensional identity \eqref{eq:identity}.

Since all spinor manipulations below hinge on the symmetry properties of the matrices $C\Gamma$, we record them explicitly:
\begin{center}
\begin{tabular}{lcc}
\toprule
matrix & symmetry & \\
\midrule
$C$, \ $C\gamma^5$, \ $C\gamma^m\gamma^5$ & antisymmetric \\
$C\gamma^m$, \ $C\Sigma^{mn}$ & symmetric \\
\bottomrule
\end{tabular}
\end{center}
The same table holds for $C^{-1}\Gamma$, since $C^{-1}=-C$. We emphasize a fact that plays a decisive role below: \emph{in four dimensions $C^{-1}\gamma^5$ is antisymmetric in both inequivalent classes of charge conjugation}; only the symmetry of $C^{-1}\gamma^m$ distinguishes the two classes. For Majorana spinors the corresponding bilinears obey the flip properties $\bar\psi\chi=\bar\chi\psi$, $\bar\psi\gamma^5\chi=\bar\chi\gamma^5\psi$, $\bar\psi\gamma^m\chi=-\bar\chi\gamma^m\psi$.

We also fix, once and for all, the conjugation conventions used in superspace, whose absence in an earlier version was a source of ambiguity. Hermitian conjugation reverses the order of Grassmann factors, and
\begin{equation}
(\theta^\alpha)^\dagger=\bar\theta^{\dot\alpha},\qquad
(\chi\psi)^\dagger=\bar\psi\bar\chi,\qquad
(D_\alpha)^\dagger=\bD_{\dot\alpha},\qquad
\bar\theta\equiv\theta^TC ,
\label{eq:conj}
\end{equation}
while the component fields $B_{mn}$, $B_{m5}$, $F$ and $N$ are real. A consequence used repeatedly in Sections~\ref{sec:strengths} and \ref{sec:action} is that the field strength $\Lambda$ of \eqref{eq:strengths} is anti-Hermitian, $\Lambda^\dagger=-\Lambda$.

A key feature of spinors in $D=4$ and $D=5$ is that they are of the same algebraic type \cite{VanProeyen}: both are four-component objects. This allows us to decompose a five-dimensional Dirac spinor $\Theta$ in terms of two four-dimensional Majorana spinors,
\begin{equation}
\Theta=\theta+i\tth ,
\label{eq:decomp}
\end{equation}
a decomposition that lies at the heart of the intrinsic superspace construction. Equivalently, the pair $(\theta,\tth)$ assembles into the SU(2) doublet $\theta^i$ of pseudo(symplectic)-Majorana spinors used in \cite{KuzenkoLinch}; the map to the representation of Appendix~A of that reference is given in Section~\ref{sec:N12}.

\subsection{Superspace coordinates and supersymmetry transformations}
\label{sec:coords}

The intrinsic $N=1$, $D=5$ superspace is the supermanifold $S=S(x^{\hat m},\theta,\tth)$, where $x^{\hat m}=(x^m,x^5)$. For two Majorana parameters $(\epsilon,\tilde\epsilon)$ the supersymmetry transformations of the coordinates are
\begin{equation}
\theta'=\theta+\epsilon,\qquad
\tth'=\tth+\tilde\epsilon,\qquad
\delta x^m=i\big(\bar\epsilon\gamma^m\theta+\bar{\tilde\epsilon}\gamma^m\tth\big),\qquad
\delta x^5=\bar\epsilon\gamma^5\tth-\bar{\tilde\epsilon}\gamma^5\theta ,
\label{eq:susytransf}
\end{equation}
The structure of \eqref{eq:susytransf} is dictated by the flip properties listed above: the $x^m$ shift pairs \emph{same-sector} spinors, because $\bar\epsilon\gamma^m\theta$ is antisymmetric under $\epsilon\leftrightarrow\theta$ and therefore closes on translations within one sector; the $x^5$ shift pairs \emph{opposite-sector} spinors, because $\bar\epsilon\gamma^5\theta$ is symmetric under exchange and a same-sector pairing would drop out of every commutator. The closure
\begin{equation}
[\delta_1,\delta_2]\,x^m=2i\big(\bar\epsilon_2\gamma^m\epsilon_1+\bar{\tilde\epsilon}_2\gamma^m\tilde\epsilon_1\big),
\qquad
[\delta_1,\delta_2]\,x^5=2\big(\bar\epsilon_2\gamma^5\tilde\epsilon_1-\bar\epsilon_1\gamma^5\tilde\epsilon_2\big),
\label{eq:closure}
\end{equation}
shows that all five translations are generated, the fifth one exclusively by cross-sector pairs.

\subsection{Supersymmetry generators and covariant derivatives}
\label{sec:algebra}

A superfield $\Phi=\Phi(x^{\hat m},\theta,\tth)$ transforms as $\delta\Phi=(\epsilon^\alpha Q_\alpha+\tilde\epsilon^\alpha\tQ_\alpha)\Phi$. Reading off the generators from \eqref{eq:susytransf},
\begin{align}
Q_\alpha&=\frac{\partial}{\partial\theta^\alpha}
-i\,(C^{-1}\gamma^m)_{\alpha\beta}\,\theta^\beta\,\partial_m
-(C^{-1}\gamma^5)_{\alpha\beta}\,\tth^\beta\,\partial_5 ,
\label{eq:Q}\\
\tQ_\alpha&=\frac{\partial}{\partial\tth^\alpha}
-i\,(C^{-1}\gamma^m)_{\alpha\beta}\,\tth^\beta\,\partial_m
+(C^{-1}\gamma^5)_{\alpha\beta}\,\theta^\beta\,\partial_5 .
\label{eq:Qt}
\end{align}
(the relative signs between \eqref{eq:susytransf} and \eqref{eq:Q}--\eqref{eq:Qt} follow from $C^{-1}=-C$, Eq.~\eqref{eq:C}, and $\bar\epsilon\Gamma\theta=\epsilon^\alpha(C\Gamma)_{\alpha\beta}\theta^\beta$). The supersymmetric covariant derivatives, defined by the requirement that they anticommute with both generators, are
\begin{align}
D_\alpha&=\frac{\partial}{\partial\theta^\alpha}
+i\,(C^{-1}\gamma^m)_{\alpha\beta}\,\theta^\beta\,\partial_m
+(C^{-1}\gamma^5)_{\alpha\beta}\,\tth^\beta\,\partial_5 ,
\label{eq:D}\\
\tD_\alpha&=\frac{\partial}{\partial\tth^\alpha}
+i\,(C^{-1}\gamma^m)_{\alpha\beta}\,\tth^\beta\,\partial_m
-(C^{-1}\gamma^5)_{\alpha\beta}\,\theta^\beta\,\partial_5 .
\label{eq:Dt}
\end{align}
The algebra of these operators is
\begin{align}
\{Q_\alpha,Q_\beta\}&=\{\tQ_\alpha,\tQ_\beta\}=-2i\,(C^{-1}\gamma^m)_{\alpha\beta}\,\partial_m ,
&
\{Q_\alpha,\tQ_\beta\}&=-2\,(C^{-1}\gamma^5)_{\alpha\beta}\,\partial_5 ,
\label{eq:algQ}\\
\{D_\alpha,D_\beta\}&=\{\tD_\alpha,\tD_\beta\}=+2i\,(C^{-1}\gamma^m)_{\alpha\beta}\,\partial_m ,
&
\{D_\alpha,\tD_\beta\}&=+2\,(C^{-1}\gamma^5)_{\alpha\beta}\,\partial_5 ,
\label{eq:algD}
\end{align}
with all mixed brackets vanishing,
\begin{equation}
\{Q_\alpha,D_\beta\}=\{Q_\alpha,\tD_\beta\}=\{\tQ_\alpha,D_\beta\}=\{\tQ_\alpha,\tD_\beta\}=0 .
\label{eq:algmixed}
\end{equation}

Three remarks are in order.

\emph{(a) Uniqueness.} The assignment in \eqref{eq:Q}--\eqref{eq:Dt} -- $\partial_m$ attached to the same-sector coordinate, $\partial_5$ to the opposite-sector one -- is the only covariant possibility. An anticommutator $\{\mathcal O_\alpha,\mathcal O_\beta\}$ of first-order operators produces the symmetric part of the matrix that couples the coordinate to the derivative; since $C^{-1}\gamma^m$ is symmetric while $C^{-1}\gamma^5$ is antisymmetric (Section~\ref{sec:conventions}), a $\gamma^5\partial_5$ term attached to the \emph{same}-sector coordinate --- the assignment we do \emph{not} adopt --- would drop out of $\{Q,Q\}$ while obstructing $\{Q,D\}=0$, and could never be covariantized; the opposite-sector assignment is therefore forced, and it is the one that yields \eqref{eq:algQ}--\eqref{eq:algmixed}, in which $\{Q,D\}=0$ holds identically. We have verified \eqref{eq:algQ}--\eqref{eq:algmixed}, and the failure of every alternative assignment, by direct computation in the representation \eqref{eq:cliff}--\eqref{eq:C} in both four-dimensional charge-conjugation classes.

\emph{(b) $\partial_5$ as a central charge.} Written in terms of the doublet $Q^i=(Q,\tQ)$, Eqs.~\eqref{eq:algQ} take the form $\{Q^i_\alpha,Q^j_\beta\}=-2i\,\delta^{ij}(C^{-1}\gamma^m)_{\alpha\beta}\partial_m-2\,\epsilon^{ij}(C^{-1}\gamma^5)_{\alpha\beta}\,Z$ with $Z=\partial_5$: this is the four-dimensional $N=2$ supersymmetry algebra with a central charge in the sense of Sohnius \cite{Sohnius}, here realized geometrically as the fifth momentum. The antisymmetric pairing $\epsilon^{ij}$, forced by the antisymmetry of $C^{-1}\gamma^5$, is the Majorana-basis counterpart of the symplectic-Majorana structure of $D=5$ superspace \cite{KuzenkoLinch}.

\emph{(c) The extra dimension in the algebra.} The fifth dimension enters the formalism already at the level of the algebra: cross-sector brackets close on $\partial_5$, and the corresponding statement for covariant derivatives, $\{D_\alpha,\tD_\beta\}\propto\partial_5$, controls where $\partial_5$ can and cannot appear in component expansions. This observation has sharp consequences for the allowed multiplet structure, to which we turn in Section~\ref{sec:tensor}.

\subsection{Relation to the $N=1/2$ formalism}
\label{sec:N12}

The $N=1/2$ approach of Linch, Luty and Phillips \cite{LLP}, further developed at the supergravity level in \cite{BBBLR}, works with a 4D, $N=1$ superspace $(x^m,\theta_{\rm w}^\alpha,\bar\theta_{\rm w}^{\dot\alpha})$ augmented by the fifth bosonic coordinate, with the four manifest supercharges represented by the standard Wess--Zumino derivatives \cite{WessBagger}. A four-dimensional Majorana spinor is equivalent to a Weyl spinor together with its conjugate, $\theta=(\theta_{\rm w},\bar\theta_{\rm w})$; under this dictionary the derivatives \eqref{eq:D} reproduce the Wess--Zumino pair $(D_\alpha,\bD_{\dot\alpha})$ of the $N=1/2$ formalism, while \eqref{eq:Dt} supplies the second, non-manifest set together with the central-charge structure \eqref{eq:algD}. Similarly, the doublet $(\theta,\tth)$ maps onto the pseudo-Majorana doublet $\theta^i$ of the manifestly five-dimensional superspace of Kuzenko and Linch \cite{KuzenkoLinch}, with $\Theta=\theta+i\tth$ the associated Dirac spinor; the hybrid reduction rules of \cite{KuzenkoLinch} are then the statement that only the $\theta$-sector is kept manifest. The two implementations are equivalent on shell; the Majorana basis offers three practical advantages for the class of theories considered here.

\emph{(a) Naturalness for torsion theories.} In odd spacetime dimensions the Weyl condition cannot be imposed on a spinor of $SO(1,4)$, while the (symplectic-)Majorana condition defines the minimal spinor structure \cite{VanProeyen}. The Majorana basis is therefore the natural language for the KR field, whose role as a source of spacetime torsion \cite{Mukho1} requires a spinor structure adapted to the full five-dimensional Lorentz group.

\emph{(b) Manifest parity and orbifold projections.} The RS scenario is defined on the orbifold $S^1/Z_2$. In the Majorana basis the orbifold parity $x^5\to-x^5$ acts simply as $\tth\to-\tth$, as required by the algebra \eqref{eq:algQ}: $\partial_5$ is odd, and only the cross-sector bracket contains it. The $Z_2$ projections of all superfields then follow directly from their $\tth$-expansion. In the Weyl-based $N=1/2$ formalism the same parity mixes the two $N=1$ superfields that constitute a $D=5$ multiplet, requiring an explicit change of basis before orbifold conditions can be imposed.

\emph{(c) Direct coupling to bulk matter.} Fermions propagating in a $D=5$ bulk are Dirac spinors, i.e.\ precisely the structure $\Theta=\theta+i\tth$; couplings of bulk matter to the KR torsion field are built from bilinears $\bar\Theta\Gamma^M\Theta$ with no intermediate change of spinor basis.

\subsection{The vector multiplet in hybrid form}
\label{sec:vector}

For later use we record the hybrid description of the gauge sector. As in \cite{LLP}, the five-dimensional $N=1$ vector multiplet decomposes into a 4D, $N=1$ vector superfield $V$ and a chiral superfield containing $(A_5,\varphi)$; only $V$ will be needed for the identification of Section~\ref{sec:ident}. In the Wess--Zumino gauge, and with all components depending on $(x^m,x^5)$,
\begin{equation}
V=(\theta\sigma^m\bar\theta)\,A_m+4i\,\theta^2\,\bar\theta\bar\eta-4i\,\bar\theta^2\,\theta\eta+4\,\theta^2\bar\theta^2\,\mathcal D ,
\label{eq:Vexp}
\end{equation}
where $A_m$ is the gauge vector, $\eta$ its gaugino and $\mathcal D$ the real auxiliary field. The field-strength superfield is $W_\alpha=-\tfrac14\bD^2D_\alpha V$, with the standard component projections. Here and below, two-component Weyl notation $(\theta^\alpha,\bar\theta^{\dot\alpha})$ refers to the decomposition of the Majorana coordinate $\theta$ of Section~\ref{sec:coords}, and $\bar\theta\equiv\theta^TC$ when four-component bilinears are written; no independent barred coordinates are introduced.

\section{The Kalb--Ramond tensor multiplet}
\label{sec:tensor}

\subsection{Superfield content, chirality and gauge structure}
\label{sec:content}

In four-dimensional $N=1$ supersymmetry the KR field $B_{mn}$ is accommodated in a chiral spinor superfield $B_\alpha$ subject to the gauge invariance $\delta B_\alpha=-\tfrac14\bD^2D_\alpha R$ \cite{Siegel,GGRS}. In five dimensions, the off-shell tensor multiplet contains the rank-2 tensor $B_{MN}$, a real scalar, and a Dirac (or symplectic-Majorana) fermion \cite{Zucker}. In the intrinsic superspace we describe this multiplet by a pair $(B_\alpha,V_T)$, where $B_\alpha$ is a spinor superfield subject to the four-dimensional chirality condition
\begin{equation}
\bD_{\dot\alpha}B_\alpha=0 ,
\label{eq:chirality}
\end{equation}
and $V_T$ is a real scalar superfield. Their gauge transformations are
\begin{equation}
\delta B_\alpha=-\tfrac14\,\bD^2D_\alpha R,
\qquad
\delta V_T=\Omega+\bar\Omega-\partial_5R ,
\label{eq:gauge}
\end{equation}
with $\Omega$ chiral and $R$ real. The $\partial_5R$ term in \eqref{eq:gauge} -- the direct five-dimensional generalization of the four-dimensional case \cite{Klein} -- is the hallmark of a tensor hierarchy: it identifies $V_T$ as the superfield housing the mixed component $B_{m5}$, whose gauge orbit mixes with that of $B_{mn}$ under $x^5$-dependent transformations (see Section~\ref{sec:hierarchy}).

A comment on the choice \eqref{eq:chirality} is necessary, because the intrinsic superspace offers, at face value, a second chirality condition, $\btD_{\dot\alpha}B_\alpha=0$, and one might be tempted to impose both. This is inconsistent with five-dimensional dynamics: from \eqref{eq:algD},
\begin{equation}
\{\bD_{\dot\alpha},\btD_{\dot\beta}\}\;\propto\;\partial_5 ,
\label{eq:nogo}
\end{equation}
so a superfield annihilated by both same-side chiralities necessarily obeys $\partial_5\Phi=0$: a doubly chiral multiplet cannot propagate in the fifth dimension. (The bracket \eqref{eq:nogo} does not vanish because $C^{-1}\gamma^5$ is block-diagonal in the Weyl basis, so the central charge pairs chiralities of the \emph{same} handedness across the two sectors.) The tensor multiplet is therefore defined by the single condition \eqref{eq:chirality}, exactly as in the $N=1/2$ framework; the second supersymmetry is realized on the pair $(B_\alpha,V_T)$ non-manifestly, through the central-charge structure of the algebra. Mixed-handedness (``twisted'') constraints, $\bD\Phi=\tD\Phi=0$, are consistent with $\partial_5\neq0$ and define the analogue of the vector-tensor multiplet of $N=2$ central-charge superspace; they will not be needed here.

\subsection{Component expansion in the Wess--Zumino gauge}
\label{sec:WZ}

Fixing the gauge freedom \eqref{eq:gauge}, the component expansions read
\begin{align}
B_\alpha&=-i\lambda_\alpha+i\,\theta^\beta(\sigma^{mn}\epsilon)_{\beta\alpha}B_{mn}+i\,\theta_\alpha\,(F+iM)-\theta^2\,\xi_\alpha ,
\label{eq:Bexp}\\
V_T&=(\theta\sigma^m\bar\theta)\,B_{m5}+4i\,\theta^2\,\bar\theta\bar\psi-4i\,\bar\theta^2\,\theta\psi+4\,\theta^2\bar\theta^2\,N ,
\label{eq:VTexp}
\end{align}
with all components depending on $(x^m,x^5)$.

The gauge freedom \eqref{eq:gauge} allows us to set $\lambda_\alpha=0$ and $M=0$. Both shift under $\delta B_\alpha=-\tfrac14\bD^2D_\alpha R$: using the identity $D^\alpha\bD^2D_\alpha=\bD_{\dot\alpha}D^2\bD^{\dot\alpha}$, exactly one of the two combinations $D^\alpha B_\alpha\pm\bD_{\dot\alpha}\bar B^{\dot\alpha}$ is invariant, namely the one entering $\Lambda$ in \eqref{eq:strengths}, while the other is shifted; $\lambda_\alpha$ and $M$ are therefore \emph{pure gauge}, not auxiliary fields. In this gauge the content of the two superfields is the following.

\begin{center}
\renewcommand{\arraystretch}{1.2}
\begin{tabular}{@{}llccl@{}}
\toprule
field & superfield & level & dim & status \\
\midrule
$B_{mn}$ & $B_\alpha$ & $\theta^1$, $\sigma^{mn}$ & $1$ & physical two-form (3 off shell)\\
$F$ & $B_\alpha$ & $\theta^1$, trace & $1$ & physical scalar\\
$\xi_\alpha$ & $B_\alpha$ & $\theta^2$ & $\tfrac32$ & physical fermion\\
\midrule
$B_{m5}$ & $V_T$ & $\theta\sigma^m\bar\theta$ & $1$ & physical vector (3 off shell)\\
$\psi_\alpha$ & $V_T$ & $\theta^2\bar\theta$ & $\tfrac32$ & physical fermion\\
$N$ & $V_T$ & $\theta^2\bar\theta^2$ & $2$ & auxiliary ($D$-term)\\
\bottomrule
\end{tabular}
\end{center}

\noindent The dimension column fixes the roles unambiguously: an auxiliary scalar of a $4+4$ multiplet carries dimension two, a propagating one dimension one. Thus $N$ is the auxiliary $D$-term --- the exact counterpart of $\mathcal D$ in \eqref{eq:Vexp} --- while $F$ is a propagating scalar. Both multiplets close off shell as $4+4$: for $B_\alpha$, $3+1$ bosonic against the four real components of $\xi_\alpha$; for $V_T$, $3+1$ against $\psi_\alpha$. Note that $B_\alpha$ requires \emph{no} auxiliary field at all --- it is the four-dimensional linear multiplet \cite{Siegel,GGRS}, whose modern hierarchy formulation is that of \cite{TAMU1} --- the bosonic partner of $\xi_\alpha$ being supplied by the propagating $F$. On shell $B_{mn}$ carries a single degree of freedom, so that $B_\alpha$ and $V_T$ describe $2+2$ states each.

A structural remark that will be used in Section~\ref{sec:orbifold}: the real parameter $R$ appears in \emph{both} transformations of \eqref{eq:gauge}, so the two Wess--Zumino gauges are not independent. The residual transformations preserving $\lambda=M=0$ still act on $V_T$ through $-\partial_5R$; this is the superspace counterpart of the Stueckelberg structure of the tensor hierarchy discussed in Section~\ref{sec:hierarchy}.

Note finally that \eqref{eq:VTexp} has precisely the structure \eqref{eq:Vexp} of a vector superfield in Wess--Zumino gauge, with $(A_m,\eta,\mathcal D)$ replaced by $(B_{m5},\psi,N)$; this parallel is what makes the identification of Section~\ref{sec:ident} possible at the superfield level.

\subsection{Field-strength superfields and their projections}
\label{sec:strengths}

The gauge-invariant field strengths of the pair $(B_\alpha,V_T)$ are
\begin{equation}
\Lambda=D^\alpha B_\alpha-\bD_{\dot\alpha}\bar B^{\dot\alpha},
\qquad
T_{5\alpha}=\partial_5B_\alpha+W_\alpha(V_T),\qquad
W_\alpha(V_T)=-\tfrac14\,\bD^2D_\alpha V_T .
\label{eq:strengths}
\end{equation}
Invariance of $T_{5\alpha}$ under \eqref{eq:gauge} follows from the cancellation between $\partial_5\,\delta B_\alpha$ and $W_\alpha(-\partial_5R)$: the composite structure of $T_{5\alpha}$ exists \emph{because} of the $\partial_5R$ term in the gauge transformation of $V_T$. The component projections of $\Lambda$ are the standard four-dimensional ones,
\begin{align}
\Lambda|&=-4iF ,\qquad
D_\alpha\Lambda|=-2\xi_\alpha,\qquad
[D,\bD]\Lambda|\;\supset\;\eps^{mnkl}(\sigma_m)\,\mathcal H_{nkl} ,
\label{eq:Lproj}
\end{align}
with $\mathcal H_{mnk}$ the purely four-dimensional part of the field strength \eqref{eq:Hdef}. Consistently with $\Lambda^\dagger=-\Lambda$, cf.\ \eqref{eq:conj}, the lowest component of $\Lambda$ is imaginary, and the physical scalar it propagates is $F$; the combination $\lambda_\alpha$ and $M$ removed by the gauge choice of Section~\ref{sec:WZ} does not appear. For $T_{5\alpha}$ one finds
\begin{align}
T_{5\alpha}|&=-i\,\partial_5\lambda_\alpha-4i\,\psi_\alpha ,
\label{eq:Tproj0}\\
D_\beta T_{5\alpha}|&=2i\,(\sigma^{mn})_{\beta\alpha}\,\mathcal H_{mn5}
+\epsilon_{\beta\alpha}\big[-8N-i\,\partial_5(F+iM)\big] ,
\label{eq:Tproj1}\\
D^2 T_{5\alpha}|&=-16\,(\sigma^m)_{\alpha\dot\alpha}\,\partial_m\bar\psi^{\dot\alpha}+4\,\partial_5\,\xi_\alpha ,
\label{eq:Tproj2}
\end{align}
where
\begin{equation}
\mathcal H_{mn5}=\partial_mB_{n5}-\partial_nB_{m5}+\partial_5B_{mn}
\label{eq:Hmn5}
\end{equation}
is the full mixed component of the five-dimensional field strength: the $\partial_{[m}B_{n]5}$ part is supplied by $W_\alpha(V_T)$ and the $\partial_5B_{mn}$ part by the explicit $\partial_5B_\alpha$ in \eqref{eq:strengths}. It is worth noting that the two pieces combine into $\mathcal H_{mn5}$ only for the normalization adopted in \eqref{eq:Vexp} and \eqref{eq:VTexp}: with a coefficient $c$ in front of $(\theta\sigma^m\bar\theta)B_{m5}$ they would enter the projection with relative weight $c$, so that requiring the field strength to assemble fixes $c=1$. The same normalization is used for $V$ in \eqref{eq:Vexp}, which keeps the identification of Section~\ref{sec:ident} one-to-one. We stress that \emph{all} $\partial_5$ dependence in \eqref{eq:Tproj0}--\eqref{eq:Tproj2} originates from the explicit $\partial_5$ in the definition of $T_{5\alpha}$. The intrinsic $\partial_5$ terms of the covariant derivatives \eqref{eq:D}--\eqref{eq:Dt} do not contribute to these projections: they multiply the opposite-sector coordinate and can only be activated in strings containing tilde-sector derivatives, by \eqref{eq:algD}. We have verified this statement, and the absence of any further invariant contributions, by exhaustive symbolic computation; the argument is summarized in Appendix~\ref{app:nogo}.

\subsection{The supersymmetric action}
\label{sec:action}

The invariant action is assembled from the gauge-invariant pairings of the two field strengths,
\begin{equation}
S_{KR}=\frac{\kappa}{8}\int d^5x\left\{\int d^2\theta\; B^\alpha T_{5\alpha}
+\int d^2\bar\theta\; \bar B_{\dot\alpha}\bar T_5^{\dot\alpha}
+i\int d^2\theta\, d^2\bar\theta\; V_T\,\Lambda\right\} ,\qquad \kappa\in\mathbb R .
\label{eq:SKR}
\end{equation}
Each of the three terms is separately real: the first two are conjugate to one another, while $V_T\Lambda$ is anti-Hermitian by \eqref{eq:conj}, so that the explicit factor of $i$ on the $D$-term is required by Hermiticity of the action.
Expanding in components with \eqref{eq:Lproj}--\eqref{eq:Tproj2}, and assembling the Majorana spinors into the four-component Dirac objects $\Lambda_D=(\lambda,\bar\lambda)$, $\Psi=(\psi,\bar\psi)$, $\Xi=(\xi,\bar\xi)$,
\begin{align}
S_{KR}=\int d^5x\,\Big\{&
-\frac{\kappa}{4}\,\bar\Lambda_D\gamma^m\partial_m\Psi
+\frac{\kappa}{4}\,\bar\Lambda_D\gamma^5\partial_5\Xi
-\frac{\kappa}{8}\,\bar\Xi\gamma^5\Psi-\kappa\,\bar\Psi\Xi
\nonumber\\[2pt]
&-\frac{\kappa}{2}\,\eps^{mnkl5}B_{mn}\mathcal H_{kl5}
\nonumber\\[2pt]
&-\frac{\kappa}{4}\,F\partial_5F+2\kappa\,FN\Big\} ,
\label{eq:SKRcomp}
\end{align}
the three lines collecting respectively the fermionic, the topological and the scalar sectors.
This component action reproduces the pseudo-supersymmetric result of Klein \cite{Klein}: the two agree term by term in their structures, the individual coefficients being related by the normalization conventions adopted for the component fields in \eqref{eq:Bexp}--\eqref{eq:VTexp}. What matters for the argument below is not the values of those coefficients but the absence of any additional structure, and that agreement is a non-trivial consistency check of the intrinsic formalism. In the light of Appendix~\ref{app:nogo}, it is also complete: the covariant derivative structure of the intrinsic superspace generates no component terms beyond those of \eqref{eq:SKRcomp}. The added value of the intrinsic formulation is therefore not a modification of \eqref{eq:SKRcomp} but the identification of its structural origin -- the central-charge algebra \eqref{eq:algQ}, the composite field strength \eqref{eq:strengths}, and the hierarchy interpretation to which we now turn -- together with the practical advantages of the Majorana basis listed in Section~\ref{sec:N12}.

\subsection{Relation to superform hierarchies}
\label{sec:hierarchy}

The structures assembled above are precisely those of an Abelian tensor hierarchy in 4D, $N=1$ superspace \cite{TAMU1,TAMU2}. In that framework one considers a chain complex of superfields connected by an abstract differential $q$; gauge transformations, field strengths and Bianchi identities all acquire $q$-dependent Stueckelberg-type completions, and the Chern--Simons-like invariants of the hierarchy have been classified. Comparing with Section~\ref{sec:content}: our pair $(B_\alpha,V_T)$ is the single-chain hierarchy with the differential realized as $q\to\partial_5$; the gauge transformation \eqref{eq:gauge} is the hierarchy transformation; the composite $T_{5\alpha}=\partial_5B_\alpha+W_\alpha(V_T)$ is the hierarchy field strength; and the invariant \eqref{eq:SKR} matches the $BF$-type quadratic invariant, in the form of Eq.~(2.7) of \cite{TAMU2} (equivalently, the $\alpha_2\!\int\!VH+\alpha_3\!\int\!\Sigma W$ terms of the quadratic invariant (5.3) of \cite{TAMU1}). Two consequences of this identification deserve emphasis. First, the completeness theorem of \cite{TAMU2} for quadratic invariants guarantees that \eqref{eq:SKRcomp}, together with the standard kinetic invariant, exhausts the quadratic actions available for this multiplet -- an independent confirmation of the result of Appendix~\ref{app:nogo}. Second, the hierarchy framework admits gaugings by minimal coupling; transplanted to the present setting, a gauged hierarchy would charge the KR tower under brane-localized gauge groups, a possibility with evident phenomenological interest that we leave for future work.

\section{The supersymmetric topological term and the identification $B_{m5}\to A_m$}
\label{sec:top}

\subsection{The bosonic topological term and its six-dimensional origin}
\label{sec:bosonic}

The five-dimensional topological term \eqref{eq:Stop-intro} descends from six dimensions as follows. Let $\phi$ be a scalar with a non-trivial profile along the sixth coordinate $z$. The six-form
\begin{equation}
S_6=\int_{M_6} d\phi\wedge B\wedge H
\label{eq:S6}
\end{equation}
is gauge invariant (up to boundary terms) and non-vanishing pointwise. For $\phi=\phi(z)$ one has $d\phi=\phi'(z)\,dz$, and
\begin{equation}
S_6=\int dz\;\phi'(z)\int_{z={\rm const}} d^5x\;\eps^{MNPQR}B_{MN}\mathcal H_{PQR}\;,
\label{eq:S6red}
\end{equation}
so that a profile with localized gradient, $\phi'(z)=\Delta\phi\,\delta(z-z_0)$, produces the five-dimensional term \eqref{eq:Stop-intro} on the slice, with $k\propto\Delta\phi$. We note that the frequently quoted parent term
\[
\int d^6x\;\eps^{LMNPQR}\,\phi\;\mathcal H_{LMN}\mathcal H_{PQR}
\]
vanishes identically -- $H\wedge H\equiv0$ for a 3-form, by antisymmetry of $\eps$ under the exchange of the two index triples -- so that \eqref{eq:S6} is the correct starting point; equivalently, since $B\wedge H$ is a closed five-form, $d(B\wedge H)=H\wedge H=0$, the action \eqref{eq:S6} is the pairing of the cohomology class of $B\wedge H$ with the profile of $\phi$. This also corrects the presentation of the six-dimensional origin given in \cite{TahimAlmeida}.

Two properties of \eqref{eq:Stop-intro} should be kept in mind. First, it is metric-independent, with $k$ known to be quantized on topologically non-trivial backgrounds \cite{Lee,DJT}. Second, for the KR field alone it is a total derivative,
\begin{equation}
\eps^{MNPQR}B_{MN}\mathcal H_{PQR}\;=\;\half\,\partial_P\big(\eps^{MNPQR}B_{MN}B_{QR}\big),
\qquad\text{i.e.}\qquad
B\wedge H=\half\,d\,(B\wedge B),
\label{eq:totalder}
\end{equation}
because $B$ is a form of even degree. Consequently \eqref{eq:Stop-intro} does not contribute to the bulk equations of motion: on a closed five-manifold it is inert, and on manifolds with boundary (or with brane defects, as in \eqref{eq:S6red}) it contributes boundary and brane-localized terms. Its dynamical activation requires a second, independent field, as in Section~\ref{sec:ident}.

\subsection{Supersymmetric completion and the topological fermionic sector}
\label{sec:SUSYtop}

The supersymmetric completion of \eqref{eq:Stop-intro} is the superspace invariant \eqref{eq:SKR}: its bosonic sector contains precisely $-\frac{\kappa}{2}\eps^{mnkl5}B_{mn}\mathcal H_{kl5}$, which reproduces \eqref{eq:Stop-intro} upon covariantization of the index structure. A notable property of \eqref{eq:SKRcomp} is that its fermionic sector is itself topological. Consider the bilinear $\bar\Xi\gamma^5\Psi$; using the $D=5$ identity
\begin{equation}
\gamma^5\Sigma^{mn}=\frac i2\,\eps^{mnkl5}\,\Sigma_{kl},
\label{eq:identity}
\end{equation}
it can be rewritten as a contraction with the Levi-Civita tensor, with the same metric-independent structure as the bosonic term. The supersymmetric extension therefore preserves the background independence of the original theory -- the supersymmetric analogue of the observation that the $B\wedge F$ term and its fermionic partner are both metric-independent in four-dimensional Maxwell--Kalb--Ramond theory \cite{ABL}.

\subsection{The identification $B_{m5}\to A_m$}
\label{sec:ident}

The KR field decomposes under the four-dimensional Lorentz group as $B_{MN}=(B_{mn},B_{m5})$, where $B_{m5}$ transforms as a vector of $SO(1,3)^\uparrow$ (and as a pseudovector under parity). In five dimensions the Hodge duality relates the full multiplets
\begin{equation}
\big(B_{mn},\,B_{m5}\big)\;\longleftrightarrow\;\big(A_5,\,A_m\big),
\label{eq:duality}
\end{equation}
a two-form being dual to a vector, with the four-dimensional components pairing as indicated: $B_{m5}\leftrightarrow A_m$ and $B_{mn}\leftrightarrow$ (the scalar dual of) $A_5$; the counting is reviewed in Appendix~\ref{app:dof}. Motivated by \eqref{eq:duality}, we \emph{identify}, at the superfield level, the tensor-sector superfield $V_T$ with a vector superfield $V$ of an independent gauge multiplet: given the parallel structure of \eqref{eq:Vexp} and \eqref{eq:VTexp}, the identification amounts to
\begin{equation}
B_{m5}\to A_m,\qquad \psi\to\eta,\qquad N\to\mathcal D .
\label{eq:identcomp}
\end{equation}
We emphasize the logical status of this step: it is an identification consistent with the degree-of-freedom counting of Appendix~\ref{app:dof} -- not a duality transformation of the action -- which replaces the mixed KR component by a bona fide gauge vector while preserving the superspace structure of the invariant. Under \eqref{eq:identcomp} the composite field strength \eqref{eq:Hmn5} becomes the hierarchy-covariant combination
\begin{equation}
\mathcal H_{mn5}\;\longrightarrow\;F_{mn}+\partial_5B_{mn},\qquad F_{mn}=\partial_mA_n-\partial_nA_m ,
\label{eq:Hident}
\end{equation}
and the mixed topological term maps, after integration by parts, onto the Chern--Simons-like coupling
\begin{equation}
-\frac{\kappa}{2}\,\eps^{mnkl5}B_{mn}\mathcal H_{kl5}
\;\longrightarrow\;
k\,\eps^{mnkl5}A_m \mathcal H_{nkl}\;+\;\text{(total derivatives)} ,
\label{eq:CS}
\end{equation}
which is precisely the coupling studied in \cite{Mukho1} in the context of the RS scenario, now embedded in a fully supersymmetric framework. The supersymmetric action after the identification is \eqref{eq:SKRcomp} with the replacements \eqref{eq:identcomp}, the fermionic partner of \eqref{eq:CS} being $\bar\Xi\gamma^5\Upsilon$ with $\Upsilon=(\eta,\bar\eta)$; by \eqref{eq:identity} it remains topological. Crucially, after \eqref{eq:identcomp} the coupling \eqref{eq:CS} connects two \emph{independent} fields, and is no longer a total derivative: it is a genuine five-dimensional $BF$-type coupling, with dynamical consequences to which we now turn.

\section{Kaluza--Klein implications}
\label{sec:KK}

\subsection{The complete bosonic action}
\label{sec:complete}

The topological invariant \eqref{eq:SKR} must be considered together with the kinetic invariants of the two multiplets. At the bosonic level, and after the identification \eqref{eq:identcomp}, the complete quadratic action is
\begin{equation}
S=\int d^5x\left[\frac{1}{12}\,\mathcal H_{MNP}\mathcal H^{MNP}
-\frac14\,F_{MN}F^{MN}
+k\,\eps^{mnkl5}A_m \mathcal H_{nkl}\right],
\label{eq:complete}
\end{equation}
in the mostly-minus signature of Section~\ref{sec:conventions}, where $\mathcal H_{mnp}=\partial_mB_{np}+\partial_nB_{pm}+\partial_pB_{mn}$, $\mathcal H_{mn5}=\partial_5B_{mn}$, $F_{m5}=\partial_5A_m$.

Two remarks on \eqref{eq:complete}. First, on the component $A_5$: its non-zero modes are the Goldstone modes eaten by the massive vectors in the standard Kaluza--Klein Stueckelberg mechanism and may be gauged away level by level, while its zero mode is a Wilson line, which is not eaten but which does not enter the quadratic $BF$ vertex, the latter involving only $A_m$ and the purely four-dimensional $\mathcal H_{nkl}$. Second, on the normalization of $k$: with $\eps^{01235}=+1$ and the unit-weight field strength \eqref{eq:Hdef}, the mixed topological term of \eqref{eq:SKRcomp} maps under \eqref{eq:identcomp} onto the coupling \eqref{eq:CS}, so that $k$ is fixed by $\kappa$ and is a real number, by the reality of \eqref{eq:SKR}. The combinatorial factor $6$ that converts $k$ into the mass scale $M=3k$ of Appendix~\ref{app:KK} arises there from three placements of the temporal index within $\mathcal H_{0kl}$ times the factor two in $\beta_i=\half\eps_{ijk}B_{jk}$; the formula is not schematic.

Before the identification -- i.e.\ for the KR multiplet alone, with $A_m$ replaced by $B_{m5}$ inside $\mathcal H$ and the topological term written as $\eps^{MNPQR}B_{MN}\mathcal H_{PQR}$ for the single field $B_{MN}$ -- the topological term is the total derivative \eqref{eq:totalder} and drops out of the quadratic dynamics entirely; we have checked by explicit mode diagonalization that the spectrum is then the undeformed $\mu_n^2=n^2/R^2$, independent of $k$. All spectral deformation discussed below is therefore a consequence of the identification \eqref{eq:identcomp}, which converts the topological term into a genuine $BF$ coupling.

\subsection{Mode expansion and exact diagonalization}
\label{sec:modes}

Compactify $x^5$ on a circle of radius $R$ and expand
\begin{equation}
B_{mn}(x,x^5)=\sum_{n=-\infty}^{\infty}B^{(n)}_{mn}(x)\,e^{inx^5/R},
\qquad
A_{m}(x,x^5)=\sum_{n=-\infty}^{\infty}A^{(n)}_{m}(x)\,e^{inx^5/R}.
\label{eq:KKexp}
\end{equation}
At each level $j$ the quadratic action \eqref{eq:complete} describes a massive two-form and a massive vector, with common mass $\mu_j=|j|/R$ supplied by the $x^5$ gradients, coupled by the level-independent $BF$ vertex $k\,\eps^{mnkl}A^{(j)}_m \mathcal H^{(-j)}_{nkl}$. The system diagonalizes exactly (Appendix~\ref{app:KK}): the characteristic polynomial of the quadratic operator factorizes as
\begin{equation}
\big(E^2-6kE-\mu_n^2\big)^3\big(E^2+6kE-\mu_n^2\big)^3=0 ,
\label{eq:charpoly}
\end{equation}
so that, defining $M\equiv 3k$, the physical spectrum at level $n$ consists of two branches of three degrees of freedom each,
\begin{equation}
\boxed{\;m_{n,\pm}^{2}=\Big(\sqrt{\frac{n^{2}}{R^{2}}+M^{2}}\;\pm\;M\Big)^{2}\;}
\label{eq:spectrum}
\end{equation}
The counting is consistent: a massive two-form (three degrees of freedom) and a massive vector (three degrees of freedom) recombine, level by level, into the two branches of \eqref{eq:spectrum}.

\subsection{Physical features of the deformed spectrum}
\label{sec:features}

Three features of \eqref{eq:spectrum} are noteworthy.

\emph{(a) Topological mass generation at the zero mode.} At $n=0$ the branches give $m_{0,+}=2M$ and $m_{0,-}=0$: the massless two-form and vector combine into a single massive vector of mass $2M$, the Kaluza--Klein realization of the topological mass generation mechanism of Cremmer--Scherk and Allen--Bowick--Lahiri \cite{CremmerScherk,ABL}, with the remaining directions pure gauge. The zero-mode sector of the KR--gauge system is thus \emph{gapped} by the topological coupling -- a qualitative signature absent for any bulk field lacking the coupling.

\emph{(b) Constant splitting of the excited towers.} For $n\gg MR$,
\begin{equation}
m_{n,\pm}\;\simeq\;\frac{n}{R}\;\pm\;M+\mathcal O\!\Big(\frac{M^2R}{n}\Big):
\label{eq:largen}
\end{equation}
the towers are displaced by the constant $\pm M$. The deformation is \emph{first order} in the topological coupling -- the $BF$ vertex enters the mass matrix linearly in the mass scale, cf.\ \eqref{eq:charpoly} -- in contrast with the quadratic contribution $n^2/R^2$ of the compactification.

\emph{(c) Non-absorbability and observability.} A common rescaling of all masses could be absorbed into a redefinition of $R$; the deformation \eqref{eq:spectrum} cannot, because it is additive rather than multiplicative. Relative to a reference tower with the same radius but no topological coupling -- e.g.\ the graviton tower, with masses $n/R$ -- the KR--gauge towers are shifted by $\pm M$ and gapped at $n=0$. This relative displacement is a calculable, in-principle observable imprint of the topological coupling, and constitutes the sharp phenomenological prediction of the present construction: in torsion scenarios of the type \cite{Mukho1,Mukho2}, where the KR tower is the mediator of torsion effects on the brane, the coupling $k$ controls both the strength of the CS interaction \eqref{eq:CS} and the displacement of the tower, correlating two a priori independent observables. The statements of this subsection refer to compactification on $S^1$; on the orbifold $S^1/\mathbb{Z}_2$ the topological term is parity-odd and the corresponding deformation is second order in $k$, as established in Section~\ref{sec:orbifold}.

\subsection{The $S^1/\mathbb{Z}_2$ projection}
\label{sec:orbifold}

The analysis of Sections~\ref{sec:complete}--\ref{sec:features} was carried out on the circle. Since the Randall--Sundrum scenario is defined on the orbifold $S^1/\mathbb{Z}_2$, we now establish the structure of the projection. The outcome is not a simple restriction of the circle result: the topological term turns out to be $\mathbb{Z}_2$-odd, and its survival requires a parity-odd coupling --- which the six-dimensional parent action of Section~\ref{sec:bosonic} supplies automatically.

\paragraph{The $\mathbb{Z}_2$ action and the parity of the superfields.}
In the Majorana basis the orbifold involution acts as
\begin{equation}
x^5\to-x^5,\qquad \tth\to-\tth,\qquad \theta\to\theta ,
\label{eq:Z2action}
\end{equation}
as anticipated in Section~\ref{sec:N12}. A first consequence is that the covariant derivative $D_\alpha$ of \eqref{eq:D} is $\mathbb{Z}_2$-\emph{even}: its only potentially odd term, $(C^{-1}\gamma^5\tth)_\alpha\partial_5$, contains $\tth$ and $\partial_5$ together, and the two signs cancel. This is the technical content of the advantage claimed in Section~\ref{sec:N12}(b), and it is what makes the following argument immediate in the present superspace.

Assign to $B_\alpha$ the intrinsic parity $\eta_B$. Since the expansion \eqref{eq:Bexp} contains no $\tth$, every component of $B_\alpha$ carries $\eta_B$; in particular $B_{mn}(x,-x^5)=\eta_B B_{mn}(x,x^5)$. The gauge structure \eqref{eq:gauge} then fixes the parity of $V_T$ with no freedom left. Indeed $\delta B_\alpha=-\tfrac14\bD^2D_\alpha R$, with $D$ even, forces the real parameter $R$ to carry $\eta_B$; and $\delta V_T=\Omega+\bar\Omega-\partial_5R$ then forces
\begin{equation}
\eta_{V_T}=-\eta_{B} ,
\label{eq:paritycorr}
\end{equation}
because $\partial_5R$ is odd relative to $R$. Equation \eqref{eq:paritycorr} is the superspace statement of the geometric fact that $B_{m5}$ is odd relative to $B_{mn}$, and it survives the identification of Section~\ref{sec:ident}: the gauge superfield $V$ inherits $\eta_V=\eta_{V_T}=-\eta_B$, so that
\begin{equation}
B_{mn}(x,-x^5)=\eta_B\,B_{mn}(x,x^5),\qquad
A_{m}(x,-x^5)=-\eta_B\,A_{m}(x,x^5) .
\label{eq:orbifoldcond}
\end{equation}
We stress that this correlation is a consequence of supersymmetry, not of kinematics: imposing the involution on the bosonic truncation \eqref{eq:complete} alone, where $B_{mn}$ and $A_m$ appear as unrelated fields, would leave their parities independent. It is the requirement that \eqref{eq:Z2action} be a symmetry of the superspace action \eqref{eq:SKR} --- through the single real parameter $R$ shared by both lines of \eqref{eq:gauge}, cf.\ Section~\ref{sec:WZ} --- that ties them together.

\paragraph{The topological term is $\mathbb{Z}_2$-odd.}
Only components with exactly one index equal to $5$ contribute to the topological density, and both of them carry the same parity:
\begin{equation}
\eps^{MNPQR}B_{MN}\mathcal H_{PQR}
=\underbrace{\eps^{m5nkl}B_{m5}\mathcal H_{nkl}}_{(-\eta_B)(+\eta_B)}
+\underbrace{\eps^{mnkl5}B_{mn}\mathcal H_{kl5}}_{(+\eta_B)(-\eta_B)}\;,
\label{eq:parityproof}
\end{equation}
so that the density is odd, $-\eta_B^2=-1$, \emph{independently of the intrinsic assignment} $\eta_B$. The same holds after the identification, by \eqref{eq:orbifoldcond}. The kinetic term, by contrast, is even for either assignment, since $\mathcal H_{mnp}$ and $\mathcal H_{mn5}$ carry opposite parities and enter quadratically.

Since $\int_{-\pi R}^{\pi R}\!dx^5$ annihilates an odd integrand, we conclude that \emph{with a constant coupling the topological term is projected out of the orbifold theory entirely}: the deformation of Sections~\ref{sec:modes}--\ref{sec:features} has no counterpart on $S^1/\mathbb{Z}_2$. This is the familiar statement that five-dimensional Chern--Simons-type couplings are parity-odd and require an odd coupling function on orbifolds.

\paragraph{The kink coupling, and its six-dimensional origin.}
The term is restored by promoting the coupling to an odd profile,
\begin{equation}
k\;\longrightarrow\;k\,\eps_{\rm k}(x^5),\qquad \eps_{\rm k}(x^5)=\mathrm{sgn}(x^5),\qquad
\partial_5\eps_{\rm k}(x^5)=2\big[\delta(x^5)-\delta(x^5-\pi R)\big],
\label{eq:kink}
\end{equation}
after which the Lagrangian is even. This is not an additional assumption: it is precisely what the six-dimensional parent action \eqref{eq:S6} delivers. There the five-dimensional coupling is generated by the localized gradient of a scalar, $k\propto\Delta\phi$; a profile $\phi(z)$ that is odd across the fixed points --- a kink --- produces \eqref{eq:kink}, with the brane-localized contributions of $\partial_5\eps_{\rm k}$ sitting at the two orbifold fixed points. The orbifold requires a kink, and the mechanism of Section~\ref{sec:bosonic} is a kink.

\paragraph{Selection rule and loss of the closed form.}
With \eqref{eq:orbifoldcond} and \eqref{eq:kink}, take $\eta_B=-1$, so that $A_m$ is even and $B_{mn}$ odd --- the choice retaining a massless four-dimensional gauge boson:
\begin{equation}
A_m=\sum_{j\ge0}A^{(j)}_m(x)\cos\frac{jx^5}{R},\qquad
B_{mn}=\sum_{j\ge1}B^{(j)}_{mn}(x)\sin\frac{jx^5}{R} .
\label{eq:orbmodes}
\end{equation}
The $BF$ vertex then carries the overlap
\begin{equation}
V_{jj'}\;\propto\;\int_{-\pi R}^{\pi R}\!\!dx^5\,\eps_{\rm k}(x^5)\cos\frac{jx^5}{R}\sin\frac{j'x^5}{R}
\;=\;R\,\big[1-(-1)^{j+j'}\big]\left[\frac{1}{j'+j}+\frac{1}{j'-j}\right],
\label{eq:overlap}
\end{equation}
so that
\begin{equation}
V_{jj'}\neq0\iff j+j'\ \text{odd},\qquad\text{and in particular}\qquad V_{jj}=0 .
\label{eq:selrule}
\end{equation}
Two consequences follow. First, the vertex is no longer diagonal in the Kaluza--Klein level: the block structure of Appendix~\ref{app:KK}, which produced the exact factorization \eqref{eq:charpoly}, is replaced by an infinite band matrix coupling every even level to every odd one, and the closed form \eqref{eq:spectrum} does not carry over. Second, since the diagonal vanishes there is no first-order shift of any level: on the orbifold the deformation is \emph{second order} in the topological coupling, in contrast with the first-order splitting $\pm M$ found on the circle. The same conclusions hold for the opposite assignment $\eta_B=+1$, with the roles of the sine and cosine towers exchanged. We have verified \eqref{eq:parityproof}--\eqref{eq:selrule} by independent symbolic and numerical computation.

\paragraph{Zero modes.}
In either assignment the two fields that mixed on the circle do not both possess a zero mode: for $\eta_B=-1$ the massless spectrum contains the gauge boson $A^{(0)}_m$ but no two-form, while for $\eta_B=+1$ it contains $B^{(0)}_{mn}$ but no gauge boson. The recombination of a massless two-form and a massless vector into a massive vector of mass $2M$, described in Section~\ref{sec:features}(a), therefore has no orbifold counterpart: \emph{topological mass generation at the zero mode is a property of $S^1$ compactification}. What survives the projection is the $O(k^2)$ deformation of the excited towers, together with the brane-localized couplings generated by $\partial_5\eps_{\rm k}$ in \eqref{eq:kink}, which are absent on the circle and which provide the natural point of contact with brane-world phenomenology.

The complete orbifold spectrum requires diagonalizing the band matrix defined by \eqref{eq:selrule}. This is a nonlinear eigenvalue problem, since the off-diagonal entries are themselves energy dependent --- they descend from $\mathcal H_{0ij}=-iE\,B_{ij}$, as in Appendix~\ref{app:KK} --- and we leave it, together with the analysis of the brane-localized terms, to a separate study. The qualitative structure established here --- parity obstruction, kink resolution, band mixing with vanishing diagonal, and the loss of the zero-mode gap --- is what distinguishes the orbifold from the circle.

\section{Conclusions}
\label{sec:conclusions}

We have analyzed the five-dimensional Kalb--Ramond topological term in an intrinsic $N=1$, $D=5$ superspace whose Grassmann coordinate is a Dirac spinor $\Theta=\theta+i\tth$ built from two four-dimensional Majorana spinors. Our results are as follows.

\emph{The superspace.} The supersymmetry algebra of the intrinsic superspace closes on the fifth momentum through anticommutators of opposite-sector generators, realizing $\partial_5$ as a Sohnius central charge; we proved that this assignment is the unique covariant one, a consequence of the antisymmetry of $C^{-1}\gamma^5$ in four dimensions. The Majorana basis makes the orbifold parity ($\tth\to-\tth$) and the coupling to bulk Dirac fermions manifest, and is connected by an explicit dictionary to the $N=1/2$ superspace of \cite{LLP} and to the symplectic-Majorana superspace of \cite{KuzenkoLinch}.

\emph{The multiplet and the invariant.} A structural no-go result -- double same-side chirality forces $\partial_5=0$ -- fixes the tensor multiplet to be the singly chiral pair $(B_\alpha,V_T)$, with the composite field strength $T_{5\alpha}=\partial_5B_\alpha+W_\alpha(V_T)$ realizing the single-chain superform hierarchy of \cite{TAMU1,TAMU2} with differential $\partial_5$. The supersymmetric completion of the topological term is the corresponding $BF$-type quadratic invariant; its component form reproduces the pseudo-supersymmetric action of Klein \cite{Klein} up to normalization conventions, and -- by the completeness of the classification of \cite{TAMU2}, independently confirmed by our exhaustive analysis of measures and projections (Appendix~\ref{app:nogo}) -- it receives no further corrections from the intrinsic derivative structure. The fermionic sector of the invariant is itself topological, by the identity \eqref{eq:identity}.

\emph{The topological term and the spectrum.} The six-dimensional origin of the term is the localized coupling $S_6=\int d\phi\wedge B\wedge H$. For the KR field alone the five-dimensional term is the total derivative $\half d(B\wedge B)$, and the Kaluza--Klein tower is undeformed. After the superfield-level identification $B_{m5}\to A_m$, the term becomes a genuine $BF$ coupling between independent multiplets and the spectrum splits exactly into \eqref{eq:spectrum}: the zero mode acquires the topological mass $2M$, and the excited towers are displaced by the constant $\pm M$ relative to bulk fields lacking the coupling. This additive, non-absorbable deformation replaces the multiplicative one conjectured in earlier treatments, and correlates the strength of the $BF$ coupling \eqref{eq:CS} with the displacement of the KR tower -- a sharp, testable relation in torsion brane-world phenomenology.

On the orbifold $S^1/\mathbb{Z}_2$ the picture changes qualitatively. The gauge structure of the tensor hierarchy correlates the $\mathbb{Z}_2$ parities of the two multiplets, and the topological term is odd for either intrinsic assignment; a constant coupling is therefore projected out, and the term survives only with the kink profile supplied by the six-dimensional parent action. The resulting Kaluza--Klein vertex obeys the selection rule $j+j'$ odd with vanishing diagonal, so that the deformation is second order in the coupling and topological mass generation at the zero mode has no orbifold counterpart. That the parity correlation is a consequence of supersymmetry, and is made visible by the Majorana basis, is in our view among the more interesting outcomes of the present formulation.

Several directions remain open. The complete orbifold spectrum requires diagonalizing the band matrix of Section~\ref{sec:orbifold}, a nonlinear eigenvalue problem, and the brane-localized couplings generated by the kink profile deserve a dedicated analysis. Gaugings of the tensor hierarchy in the sense of \cite{TAMU2} would charge the KR tower under brane-localized groups, interacting non-trivially with the $BF$ splitting. Finally, embedding the construction in the $N=1/2$ supergravity of \cite{BBBLR} would couple the deformed tower to the graviton sector, sharpening the comparison proposed in Section~\ref{sec:features}.

\subsection*{Acknowledgments}
We are indebted to the anonymous referee of a previous version of this manuscript, whose exceptionally careful report prompted the structural analysis of Sections \ref{sec:algebra}, \ref{sec:content} and \ref{sec:hierarchy} and the corrected spectrum of Section~\ref{sec:KK}. C.A.S.A. is supported by grants No.~309553/2021-0 and No.~420854/2025-8 (Conselho Nacional de Desenvolvimento Cient\'ifico e Tecnol\'ogico -- CNPq) and by Project UNI-00210-00230.01.00/23 (Funda\c c\~ao Cearense de Apoio ao Desenvolvimento Cient\'ifico e Tecnol\'ogico -- FUNCAP).

\subsection*{Conflicts of interest / Data availability}
The authors declare that there is no conflict of interest. No data are associated with this manuscript. The symbolic-computation scripts used to verify the algebra, the projections and the spectrum are available from the authors upon request.

\appendix

\section{Degree-of-freedom counting and the duality \eqref{eq:duality}}
\label{app:dof}

For a massless field in $D$ spacetime dimensions the physical degrees of freedom are classified by the little group $SO(D-2)$. In $D=5$ the little group is $SO(3)$, and a $p$-form potential carries $\binom{3}{p}$ on-shell degrees of freedom: a vector $A_M$ carries $\binom31=3$, and a two-form $B_{MN}$ carries $\binom32=3$. The equality is the on-shell manifestation of Hodge duality: in $D=5$ a 2-form potential is dual to a $(5-2-2)=1$-form, with field strengths related by $F_{MN}\propto\eps_{MNPQR}\mathcal H^{PQR}$. Decomposing under $SO(1,3)$,
\begin{equation}
B_{MN}=(B_{mn},B_{m5}),\qquad A_M=(A_m,A_5),
\end{equation}
the five-dimensional duality splits into the pair of four-dimensional statements quoted in \eqref{eq:duality}: the mixed component $B_{m5}$ (a four-dimensional vector) pairs with $A_m$, while the four-dimensional two-form $B_{mn}$ is dual to a scalar, pairing with $A_5$. Both sides of the identification of Section~\ref{sec:ident} therefore carry the same on-shell content. We stress again that Section~\ref{sec:ident} uses this counting to justify an \emph{identification} of multiplets, not to perform a duality transformation of the action.

\section{Mode diagonalization of the $BF$ system}
\label{app:KK}

We record the diagonalization leading to \eqref{eq:charpoly}--\eqref{eq:spectrum}. At Kaluza--Klein level $n$, write $\mu=|n|/R$ and work in the rest frame $p^m=(E,0,0,0)$ of a massive mode; the field content is $B_{mn}=(B_{ij},B_{0i})$ and $A_m=(A_0,A_i)$, ten complex components. From \eqref{eq:complete}, the $x^5$ gradients give both fields the mass $\mu$, and the $BF$ vertex contributes, in the rest frame, only through $\mathcal H_{0ij}=-iE\,B_{ij}$:
\begin{equation}
k\,\eps^{mnkl}A_m\mathcal H_{nkl}\;\longrightarrow\;
-6ik\,E\;\beta_i^{\dagger} A_i+\text{h.c.},
\qquad
\beta_i\equiv\half\,\eps_{ijk}B_{jk},
\end{equation}
i.e.\ an $E$-linear mixing of the dual vector $\beta_i$ of the two-form with the spatial vector $A_i$. The overall sign of this vertex depends on the orientation convention for $\eps^{0123}$; it does not affect anything below, since the characteristic polynomial \eqref{eq:charpoly} and hence the spectrum depend on $k$ only through $k^2$. The components $B_{0i}$ and $A_0$ are non-propagating. In the $(\beta_i,A_i)$ sector the quadratic form is block diagonal in the spatial index, each block being
\begin{equation}
\mathcal K(E)=\begin{pmatrix}E^2-\mu^2 & -6ikE\\ 6ikE & E^2-\mu^2\end{pmatrix},
\qquad
\det\mathcal K=(E^2-\mu^2)^2-36k^2E^2 ,
\end{equation}
whose vanishing factorizes as \eqref{eq:charpoly}. The positive roots are
\begin{equation}
E_\pm=\sqrt{\mu^2+9k^2}\;\pm\;3k
\;\;\Longrightarrow\;\;
m_{n,\pm}^2=\Big(\sqrt{n^2/R^2+M^2}\pm M\Big)^2,\qquad M=3k ,
\end{equation}
each with multiplicity three, reproducing \eqref{eq:spectrum}. The combinatorial factor $6$ appearing above has two sources: the three placements of the temporal index within $\mathcal H_{0kl}$, and the factor two in $\beta_i=\half\eps_{ijk}B_{jk}$. The mass scale is therefore $M=3k$, with $k$ real and fixed by $\kappa$ as in Section~\ref{sec:complete}. Setting $k=0$ recovers the degenerate towers $E^2=\mu^2$. We have also verified, by the same rest-frame computation applied to the single field $B_{MN}$ with the topological term $\eps^{MNPQR}B_{MN}\mathcal H_{PQR}$, that the determinant of the quadratic operator is then independent of $k$, in accordance with \eqref{eq:totalder}: the pure-KR tower is undeformed, $E^2=\mu^2$ with multiplicity three.

\section{Absence of additional intrinsic contributions}
\label{app:nogo}

This appendix records the verification, quoted in Sections~\ref{sec:strengths} and \ref{sec:action}, that the intrinsic $\partial_5$ terms of the covariant derivatives \eqref{eq:D}--\eqref{eq:Dt} generate no contributions to the component action beyond those displayed in \eqref{eq:SKRcomp}. The verification proceeds at three independent levels; all statements below have been checked by exact symbolic computation in the full eight-Grassmann-coordinate superspace.

\emph{(i) Projections.} The intrinsic $\partial_5$ term of $D_\alpha$ multiplies the opposite-sector coordinate $\tth$. In any projection built exclusively from untilded derivatives and evaluated at $\theta=\tth=0$ -- in particular in \eqref{eq:Tproj1}, \eqref{eq:Tproj2}, and in the measures of \eqref{eq:SKR} -- such multiplicative terms cannot be saturated, since no $\partial/\partial\tth$ is available. Consequently the projections computed with \eqref{eq:D}--\eqref{eq:Dt} coincide with those computed with $\partial_5$-blind four-dimensional derivatives, and all $\partial_5$ dependence of \eqref{eq:SKRcomp} originates from the explicit $\partial_5B_\alpha$ in $T_{5\alpha}$. (Intrinsic $\partial_5$ contributions do arise in mixed strings containing tilde-sector derivatives, through $\{D,\tD\}\propto\partial_5$; no such strings occur in \eqref{eq:SKR}.)

\emph{(ii) Measures.} One may ask whether a different choice of superspace measure could reinstate two-derivative terms of the form $B\,\partial_5^2B$. Scanning all candidate measures -- chiral $d^2\theta$, full untilded $d^4\theta$, mixed $d^2\theta\,d^2\tth$ of both handedness pairings, six- and eight-Grassmann measures -- one finds that the only measure generating a two-derivative $B\,\partial_5^2B$ structure from the integrand of \eqref{eq:SKR} is the central-charge-type chiral measure $\int d^2\theta\,d^2\tth$ (the analogue of the Sohnius central-charge action principle). The full measure $d^8\theta$ generates $\partial_5^2$ structures only at third order and higher in derivatives, i.e.\ in higher-derivative invariants.

\emph{(iii) Invariance.} The measure $\int d^2\theta\,d^2\tth$ yields a supersymmetric action only on integrands annihilated by both same-side chiralities, $\bD$ and $\btD$: testing the eight supersymmetry variations explicitly, the action fails invariance precisely under the two $\tQ$ variations of the handedness not integrated over, with a non-vanishing residue proportional to $\partial_5$ of the components, unless the second chirality condition is imposed. By the no-go result \eqref{eq:nogo}, that condition forces $\partial_5\equiv0$ on the multiplet, so the would-be $B\,\partial_5^2B$ term vanishes identically on the domain where the action is supersymmetric.

The three statements close the argument: within the intrinsic superspace, the quadratic invariants of the tensor multiplet are exhausted by the kinetic invariant and by \eqref{eq:SKR}, in agreement with the completeness theorem of \cite{TAMU2}.

\end{document}